\documentclass[prb,showpacs,showkeys,preprintnumbers,amsmath,amssymb,twocolumn]{revtex4-1}
\usepackage[T1]{fontenc} 
\usepackage{makeidx} 
\usepackage{graphicx} 
\usepackage{dcolumn} 
\usepackage{array} 
\usepackage{amssymb} 
\usepackage{amsmath}
\usepackage{textcomp}
\usepackage{rotating}
\usepackage{wasysym}
\usepackage{multirow}
\usepackage{subfigure}
\usepackage{eucal}
\usepackage{mathrsfs}
\usepackage{units}
\usepackage{rotating}
\usepackage[all]{xy}
\usepackage{float} 
\usepackage{amsmath} 
\usepackage{amsfonts} 
\usepackage{bm} 
\usepackage[amssymb]{SIunits}

\begin{document}

\title{Step-like features on caloric effects of graphenes}

\author{M.S. Reis}\email{marior@if.uff.br}\affiliation{Instituto de F\'{i}sica, Universidade Federal Fluminense, Av. Gal. Milton Tavares de Souza s/n, 24210-346, Niter\'{o}i-RJ, Brasil}

\date{\today}

\begin{abstract}
We considered a graphene nano-ribbon with a longitudinal electric field (along $x$ direction) and a transversal magnetic field (along $z$ direction), and then observe (i) the electrocaloric effect ruled by an applied magnetic field and (ii) the magnetocaloric effect ruled by an applied electric field. We focused our attention to the limit of low temperatures, and then observed interesting step-like features. For each filled Landau level $n$, created by the applied magnetic field, both caloric effects increase proportionally to $n+1/2$; and this step measures either important graphene properties (like Fermi velocity) or quantum fundamental quantities (like Planck constant and magnetic flux quantum).
\end{abstract}

\maketitle

\section{Introduction}

Caloric effects, like magneto (MCE) and electro (ECE), are intrinsic and exciting properties of magnetic and electric materials, in which they are able to exchange heat ($\Delta Q=T\Delta S$) with a thermal reservoir, under a field change ($\Delta B$ or $\Delta E$). This amount of heat is directly related to the entropy change $\Delta S(T,\Delta X)=S(T,X)-S(T,0)$, usually used to characterize the caloric properties of materials. $X$ is a field - either magnetic $B$ or electric $E$; or even pressure $p$, considering the barocaloric effect\cite{reis2013fundamentals}. The cooling devices based on these effects are promising candidates to substitute, in a near feature, the common devices, i.e., those based on compression-expansion of Freon-like gases (for instance, air conditioners and household refrigerators). This probable scenario arises since cooling devices fit in a clean, safe and sustainable technology; where these systems do not use any noxious gases to the ozone layer and, in addition, have greater efficiency of cooling power and lower energy consumption when compared with the traditional devices\cite{tishin_book}. However, applications of these effects are not limited to room temperature, and the best example is the adiabatic demagnetization refrigerator, that can reach mK scale \cite{Timbie1990271}.

Thus, it is simple to understand why the scientific community has never explored materials until their limits and never observed step-like features on caloric effects, since most of the research world wide has been devoted to practical purposes on magnetic-, electric- and baro-refrigeration\cite{Valant_2012_980,Scott_2011_229,AdvMat_21_2009_1983,JAP_109_2011_53515}; with a huge amount of experimental work, in what concerns materials science, and, on the other hand, only a few mean-field theoretical models \cite{pedroreport}. In addition, these caloric effects are maximized around the critical temperature of the materials and therefore ferro (magnetic and electric) materials are the most focused by the community\cite{Science_311_2006_1270,Science_321_2008_821,APL_100_2012_192902,PRL_108_2012_167604, tishin_book}, as well as those with other kind of ordering and correlations, like magneto/electric-structural dependencies\cite{tishin_book}.

In opposition to this tendency, recently, diamagnetic materials received due attention never received before \cite{MCE_OSC_DT, MCE_OSC_DS,reisa2013oscillating}, and an oscillatory behavior, due to the crossing of the Landau levels through the Fermi level \cite{greiner}, was predicted, in analogy to the de Haas-van Alphen effect. More recently, we extended our analysis to a graphene (a planar sheet of Carbon atoms packed in a honeycomb lattice), with further investigation on the oscillating magnetocaloric\cite{mce_grafeno} and electrocaloric\cite{ece_grafeno} effects on this material. Remarkable, these oscillations occur at c.a. 1 K for non-relativistic 3D diamagnetic material\cite{MCE_OSC_DT, MCE_OSC_DS,reisa2013oscillating}, while for graphenes it occurs at c.a. 100 K, due to its huge Fermi velocity\cite{mce_grafeno}. Generally speaking, anomalous phenomena on graphenes are ruled by the remarkable relativistic-like spectrum of electrons and holes, and one interesting effect experimentally verified is the abnormality of the QHE\cite{nature1,nature2,prl1,poincare} - with a  quantization condition shifted by a half-integer. 

Another important feature of graphene is the de Haas-van Alphen effect (dHvA), in which the magnetization oscillates periodically in a sawtooth pattern as a function of reciprocal magnetic field $1/B$, with period given by\cite{prb1,JPCM_22_2010_115302, PLA_375_2011_3624}:
\begin{equation}\label{bm1}
\Delta\left(\frac{1}{B}\right)=\frac{1}{B_{n+1}}-\frac{1}{B_n}=\frac{2}{\phi_0 N_0}
\end{equation}
Above, $B_{n+1}$ and $B_n$ are the magnetic induction intensity corresponding to two neighboring Landau levels $n$ which cross the Fermi level in succession; and $\phi_0=\pi\hbar/e=2.06\times10^{-15}$ Tm$^2$ is the magnetic flux quantum. 

Thus, the present effort further analysis the caloric effect on graphenes: observes and explores the step-like features on the entropy change. 

\section{Brief survey on caloric effects}

Electronic properties of graphenes are quite different from those of non-relativistic 2D (and even 3D) diamagnetic materials, since electrons in graphene have zero effective mass and these behave like relativistic particles, described by the Dirac equation\cite{livro_grafeno}. The energy spectra for a nano-ribbon graphene sheet on $x-y$ plan with $\vec{E}=(-E,0,0)$ and $\vec{B}=(0,0,B)$, i.e., electric field along the graphene plan and magnetic field perpendicular to graphene plan, is\cite{prl_referee,JPCM_22_2010_115302, PLA_375_2011_3624}:
\begin{equation}
E_n=\hbar\omega^\prime\sqrt{n}+\hbar v_F\beta k_y
\end{equation}
where $n=0,1,2,\cdots$ represents the Landau level index, 
\begin{equation}
\hbar\omega^\prime=\sqrt{2\hbar eB}v_F(1-\beta^2)^{3/4}
\end{equation}
and
\begin{equation}\label{beta}
\beta=\frac{E}{v_FB}<1,
\end{equation}
Above, $k_y=2\pi l/L_y$ ($l=0,\pm1,\pm2$,...) corresponds to the translational symmetry along the $y$ axis and is related to the size ($L_y$)  of the graphene along $y$ direction. Finally, $v_F=10^6$ m/s stands for the Fermi velocity.

This system was explored considering the magnetocaloric effect with a longitudinal applied electric field, and the entropy change is\cite{graphene_MCE_electric}:
\begin{equation}\label{MCEFINAL}
\Delta S(T,  \Delta B, E)=S_{cos}(T,B,E)+S_{per}(T,B,E)
\end{equation}
More recently, the electrocaloric effect with a transversal applied magnetic field was also explored and, similarly to above, the entropy change is\cite{ece_grafeno}:
\begin{equation}\label{ECEfinal}
\Delta S(T, B, \Delta  E)=S_{cos}(T,B,E)+S_{per}(T,B,E)-S_{cos}(T,B,0)
\end{equation}
Note we are describing the ECE in a general context, as an entropy change due to an electric field change, even graphene being a conductor. This entropy change is then related to the change on the Landau structure due to an applied electric field. In equations \ref{MCEFINAL} and \ref{ECEfinal}, those terms are (i) an oscillatory contribution to the entropy\cite{graphene_MCE_electric}:
\begin{equation}\label{cos}
S_{cos}(T,B,E)=2 k_B\frac{N_0}{m}\left(1-\beta^2\right)^{3/4}\cos(\pi m)\mathcal{T}(x)
\end{equation}
and, (ii) a periodic one, due to the additional energy induced by the electric field on the level partially occupied\cite{graphene_MCE_electric}:
\begin{equation}\label{sth}
S_{per}(T,B,E)=2\pi^2 k_BL_x\beta A_m\frac{N_0}{m}\sqrt{N_0\pi}\mathcal{T}(x)
\end{equation}
 Above,
\begin{equation}\label{tau}
\mathcal{T}(x)=\frac{xL(x)}{\sinh(x)}\;\;\;\textnormal{and}\;\;\;L(x)=\coth(x)-\frac{1}{x}
\end{equation}
is the Langevin function. In addition,
\begin{equation}\label{xk}
x=\frac{\phi_0}{B}\frac{k_BT}{\tilde{v}_F}\frac{1}{(1-\beta^2)^{3/4}},
\end{equation}
$N_0=10^{16}$ m$^{-2}$ is the density of charge carriers\cite{NN,JAP},
\begin{equation}\label{mmesmo}
m=N_0\frac{\phi_0}{B}
\end{equation}
and $\tilde{v}_F=\hbar v_F/2\pi\sqrt{N_0\pi}=9.43\times10^{-38}$Jm$^2$. Note both quantities, $m$ and $x$, are dimensionless. Finally, 
\begin{equation}
A_m=\left[m^2-2m(2\sigma+1)+4\sigma(\sigma+1)\right]
\end{equation}
where $\sigma=\lfloor m/2\rfloor$ (the $\lfloor\;\rfloor$ symbol mean the $floor$ function, i.e., the integer part of the argument). The periodic behavior of $A_m$ can be seen in reference \onlinecite{graphene_MCE_electric}. It is important to note that, for odd values of $m$, this function is $-1$ and, on the other hand, it assumes $0$ for even values of $m$. For further understanding on the above contributions to the magneto- and electro-caloric effects of graphenes, see references \onlinecite{mce_grafeno,graphene_MCE_electric,ece_grafeno}.

\section{Step-like behavior}
To observe step-like features on those caloric effects, two approximations must be done: (i) $\beta\ll 1$ - retaining only linear contributions on $\beta$. For practical purposes, considering $E=10^6$ V/m then $m< 20$. (ii) $x\rightarrow 0$ and then $\mathcal{T}(x)\rightarrow x/3$. It means $T\ll N_0\tilde{v}_F/mk_B\approx 10$ K, considering $m=10$.

\subsection{Electrocaloric effect}

It is simple to observe that the entropy change of equation \ref{ECEfinal}, on the limits imposed before, resumes as $\Delta S(T,B,\Delta E)=S_{per}(T,B,E)$, i.e., depends only on the additional energy induced by the electric field on the level partly occupied. Thus, the entropy change can be re-written as:
\begin{equation}\label{treze}
\Delta S(T,B,\Delta E)=\frac{\Lambda}{2}mA_m
\end{equation}
where 
\begin{equation}\label{lambda}
\Lambda=\left(\frac{8}{3}\pi^4k_B^2TV_x\right)\left(\frac{1}{v_F^2}\right)\left(\frac{1}{\hbar\phi_0}\right)
\end{equation}
and $V_x=EL_x$ is the voltage applied along the $x$ direction of the graphene. Considering $L_x=10^{-8}$ m (see reference \onlinecite{PLA_375_2011_3624}), the voltage applied is of the order of 10 mV. Note the first parenthesis of the above equation depends only on the external parameters and constants; the second parenthesis depends on intrinsic graphene properties and, finally, the last one depends on quantum fundamental quantities.

\begin{figure}
\center
\includegraphics[width=8cm]{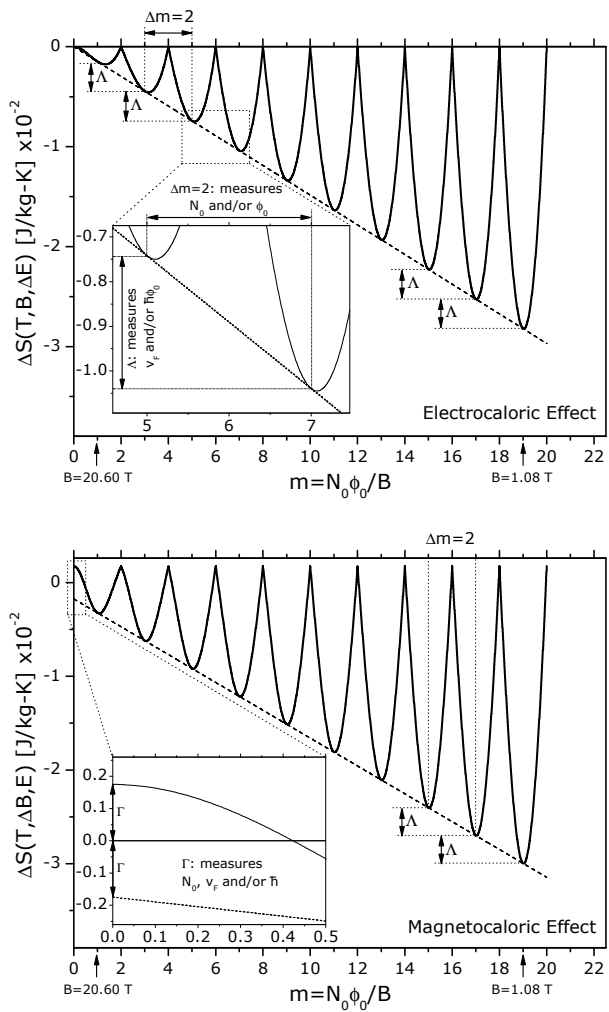}
\caption{(top) Electrocaloric effect, ruled by an applied magnetic field $B$ and (bottom) magnetocaloric effect, ruled by an applied electric field. For each filled Landau level, these caloric effects increase in $\Lambda$ (eq. \ref{lambda}). Insets: evidence that these effects are able to measure either important properties of graphenes (namely Fermi velocity $v_F$ and $N_0$) or quantum fundamental quantities (namely Planck constant $\hbar$ and magnetic flux quantum $\phi_0$). We considered $T=1$ K and 0.77 mg/m$^2$ for the graphene density. See text for further details.\label{fig}}
\end{figure}

The electrocaloric effect on equation \ref{treze} is proportional to $A_m$, that vanishes for even values of $m$ and assumes $-1$ for odd values of $m$. It has a consequence on the caloric effect, as can be seen on figure \ref{fig}-top: a periodic behavior is found and the value of $\Delta S(T,B,\Delta E)$ at each odd values of $m$, i.e., $m=2n+1$ (where $n=0,1,2,3,\cdots$) assumes
 \begin{equation}\label{qtum}
 \Delta S(T,B_n,\Delta E)=-\Lambda\left(n+\frac{1}{2}\right)
 \end{equation}
 where
\begin{equation}\label{bn}
B_n=\frac{N_0\phi_0}{2n+1}\;\;\;\Rightarrow\;\;\;\Delta\left(\frac{1}{B}\right)=\frac{2}{\phi_0 N_0}
\end{equation}
The above $\Delta(1/B)$ leads to the same result of equation \ref{bm1} and therefore we conclude that $n$ (on equation \ref{qtum}) is the Landau level index. In other words, one period of oscillation of the caloric effect is related to one Landau level.
The $\Delta S(T,B_n,\Delta E)$ difference between consecutive $n$ values, i.e., $|\Box S|=|\Delta S(T,B_{n+1},\Delta E)-\Delta S(T,B_n,\Delta E)|$ is $\Lambda=2.97 \times 10^{-3}$ J/kg-K, considering $T=1$ K.

From the experimental point of view, these results are able to measure either fundamental quantities (magnetic flux quantum $\phi_0$ and Planck constant $\hbar$) or important graphene quantities (Fermi velocity $v_F$ and density of charge carries $N_0$). Connecting those peaks with a straight line (it is easy to see that it is $-\Lambda N_0\phi_0/2B$), the abscissa axis measures, from the period of the oscillations, either $N_0$ or $\phi_0$, depending if the abscissa is expressed as $\phi_o/B$ or $N_0/B$, respectively. The ordinate axis measures $\Lambda$, since the entropy change increases at exactly this value for each consecutive $B_n$; and thus, from equation \ref{lambda}, either the Fermi velocity $v_F$ or the Planck constant $\hbar$ can be measured (remember, $\phi_0$ can be measured from the abscissa). See the inset of figure \ref{fig}-top for further details. Also note, for this limit of low temperatures, the maximum magnetic entropy change increases by decreasing the magnetic field, i.e., increasing $m$; a contrary tendency to what is observed in ordered materials\cite{reis2013fundamentals,tishin_book}.

\subsection{Magnetocaloric effect}
In what concerns the magnetocaloric effect, it is possible to obtain, analogously to before, the entropy change for the limits imposed above ($T\ll N_0\tilde{v}_F/mk_B$ and $\beta\ll 1$). Meanwhile, the magnetocaloric effect under a longitudinal applied electric field has some singularities that must be considered. Usually, the caloric effect is due to a field change, from 0 up to a finite value of field; however, it is not possible for the present case, otherwise zero magnetic field would violates the condition impose to the model ($\beta<1$ - see equation \ref{beta}). For a critical value $\beta=1$ the entire Landau structure collapses\cite{prl_referee}. Thus, here, the magnetic field change is from $B_{min}=E/v_F$ up to $B$. The result presented in equation \ref{MCEFINAL} already takes it into account. Thus, the magnetic entropy change at those limits resumes as:
\begin{equation}
\Delta S(T, \Delta B, E)=\Gamma \cos(\pi m)+\frac{\Lambda}{2}mA_m
\end{equation}
where 
\begin{equation}
\Gamma=\left(\frac{4}{3}\pi^{3/2} k_B^2T\right)\left(\frac{\sqrt{N_0}}{v_F}\right)\left(\frac{1}{\hbar}\right)
\end{equation}
As before (equation \ref{lambda}), the above equation is related to external parameters, graphene properties and quantum fundamental quantities. Considering $T=1$ K, the above constant is $\Gamma=1.75 \times 10^{-3}$ J/kg-K. The above result is presented in figure \ref{fig}-bottom and it is similar to the electrocaloric effect, presented in figure \ref{fig}-top; however, the cosine term changes mainly the behavior for high values of magnetic field (small values of $m$), and, in addition, promotes a shift on the entire curve. 

For each complete Landau level $n$, the magnetic entropy change, for $\Delta B_n: B_{min}\rightarrow B_{n}$, resumes as:
 \begin{equation}
 \Delta S(T, \Delta B_n, E)=-\Lambda\left(n+\frac{1}{2}\right)-\Gamma
 \end{equation}
It has a behavior similar to the electrocaloric case and, in addition, the entropy change increases, for each filled Landau level, in the same quantity $|\Box S|=\Lambda$.

From the experimental point of view, connecting the MCE peaks with a straight line (it is also easy to see that this line is $-\Gamma-\Lambda N_0\phi_0/2B$), the same quantities can be measured (as discussed before, in the context of the electrocaloric effect - see fig\ref{fig}-top): $N_0$ or $\phi_0$ and $v_F$ or $\hbar$. In addition, this effect allows to measure $\Gamma$, simply from the extrapolation of the straight line to high values of magnetic field (or, easier, $m\rightarrow 0$). Thus, measuring $\Gamma$, either the Fermi velocity $v_F$ or the Planck constant $\hbar$ can be obtained (remember, $N_0$ can be obtained from the abscissa).


\section{Conclusions}

We considered a graphene nano-ribbon with a longitudinal electric field (along $x$ direction) and a transversal magnetic field (along $z$ direction), and then evaluated (i) the electrocaloric effect ruled by an applied magnetic field and (ii) the magnetocaloric effect ruled by an applied electric field. We focused our attention to the limit of low temperatures, and then observed interesting features. More precisely, for each filled Landau level $n$, created by the applied magnetic field, both caloric effects increase proportionally to $n+1/2$, and steps in a exact quantity $\Lambda$ (equation \ref{lambda}), that measures either important properties of the graphenes (like the Fermi velocity $v_F$), or quantum fundamental quantities (like the Planck constant $\hbar$ and the magnetic flux quantum $\phi_0$). These results are of broad interest, since these caloric effects can be associated to intrinsic properties of graphenes and quantum fundamental quantities.
\begin{acknowledgements}
We acknowledge FAPERJ, CAPES, CNPq and PROPPI-UFF for financial support. We are in debt with Prof. P.J. von Ranke, due to his fruitful comments.
\end{acknowledgements}

\end{document}